\documentclass[aps,prl,twocolumn,floats]{revtex4-2}

\usepackage{amssymb}
\usepackage{amsmath}
\usepackage{graphicx}
\usepackage{bm}
\usepackage{dsfont}
\usepackage{mdframed}
\usepackage{color}
\usepackage{gensymb}
\usepackage{comment}
\usepackage{lipsum}
\usepackage{upgreek}
\usepackage[colorlinks=true]{hyperref}

\begin{document}

\bibliographystyle{prsty}

\title{Non-Abelian Gauge Theory for Magnons in Topologically Textured Frustrated Magnets}

\author{Ricardo Zarzuela$^{1}$ and Se Kwon Kim$^{2}$
}

\affiliation{$^{1}$ Institut f\"{u}r Physik, Johannes Gutenberg Universit\"{a}t Mainz, D-55099 Mainz, Germany\\ $^{2}$ Department of Physics, Korean Advanced Institute of Science and Technology, Daejeon 34141, South Korea}

\begin{abstract}
We develop an effective gauge theory for three-flavored magnons in frustrated magnets hosting topological textures with the aid of the quaternion representation of the SO(3) order parameter. We find that the effect of topological solitons on magnons is captured generally by the non-Abelian emergent electromagnetic fields, distinct from the previously established gauge theory for magnons in collinear magnets where the gauge theory is often restricted to be Abelian. As concrete examples, $4\pi$-vortices in two-/three-dimensional magnets and Shankar skyrmions in three-dimensional magnets are discussed in detail, which are shown to induce, respectively, the Abelian and the non-Abelian topological magnetic field on magnons, and thereby engender the topological Hall transport in textured frustrated magnets. Our work is applicable to a broad class of magnetic materials whose low-energy manifold is described by the SO(3) order parameter. We envision that the discovery of the non-Abelian magnonic gauge theory will enrich the field of magnonics as well as prompt the study of magnon transport in textured frustrated magnets.
\end{abstract}

\maketitle

\textit{Introduction.}|Magnons, quanta of spin waves, have gained significant interest over the last decade as potential building blocks for the transmission and storage of information in magnetically ordered platforms, with applications ranging from quantum information based upon magnon Bose-Einstein condensates \cite{Andrianov-PRA2014,Mohseni-NCP2022} to neuromorphic networks and stochastic/reservoir computing \cite{Barman-JPCM2021,Wang-2023}, to name a few. Remarkably, they also interact with topological solitons emerging in the collinear magnetic order, exhibiting topological properties in their spectrum and transport~\cite{Kim-PRL2016,Owerre-JPCM2016,Cheng-PRL2016,Zyuzin-PRL2016}. This interaction has been commonly described via the emergence of an Abelian electromagnetic field in the spin-space geometry for magnons \cite{Schutte-PRB2014,Tatara-PE2019, Weber-Science2022}, where magnons of different flavors (e.g., up- or down-spin in the local spin frame) are not mixed with each other, leading to the prediction of the topological magnon (spin) Hall effect in (anti)ferromagnets~\cite{vanHogdalem-PRB2013, Garst-JPD2017, Oh-PRB2015, Diaz-PRL2019}. There is an ongoing research endeavor to identify novel interactions between magnons and topological textures in magnonics.

On the other hand, the recent experimental observation of striking spin transport phenomena in glassy antiferromagnets \cite{Lair-NMat2020,Little-NMat2020,Doyle-NPhys2021} has triggered a revival of interest in frustrated magnets, namely magnetic systems with frustrated interactions dominated by isotropic exchange. It is well known that the low-energy long-wavelength description of this class of noncollinear magnetic materials is provided by the O(4) nonlinear $\sigma$-model~\cite{Dombre-PR1989,Azaria-PRL1992,Chubukov-PRL1994},
\begin{equation}
\label{eq:NsM_O4}
\mathcal{S}[\hat{R}]=\int dt\,d\vec{r}\,\frac{1}{16}\textrm{Tr}\left[\partial_{\mu}\hat{R}^{\top}\partial^{\mu}\hat{R}\right],
\end{equation}
where $\hat{R}$ denotes a three-dimensional rotation matrix, i.e., the SO(3) order parameter, $\mu=t,x,y,z$ runs over spatiotemporal indices, and we have introduced the Minkowski metric $\eta_{\mu\nu}=\textrm{diag}(1,-1,-1,-1)$. These frustrated magnets serve as a rich platform to look for unconventional thermal excitations, since the exotic SO(3) order emerging at the mesoscale presents \emph{three} fluctuation modes~\cite{Zarzuela-PRB2021}, distinct from the collinear magnets with only one or two fluctuations modes. How topological solitons unique to frustrated magnets, e.g., Shankar skyrmions and $4\pi$ vortices, interact with these SO(3) magnons has not been discussed yet. In particular, whether the emergent gauge field describing the interaction between the topological solitons and magnons in frustrated magnets is Abelian or non-Abelian and how they manifest in transport properties are open questions.

In this Letter, we address these open questions by developing a relativistic framework for the low-energy order-parameter fluctuations of frustrated magnets. More specifically, we build an effective gauge theory for SO(3) magnons that incorporates their interaction with generic topological solitons appearing in the ground state, which is reminiscent of quantum chromodynamics (QCD).  We find that the nature of the emergent gauge field depends on the class of topological solitons considered: it is Abelian for $4\pi$ vortices, non-Abelian for Shankar skyrmions and vanishing for magnetic disclinations. We discuss as well the magnon transport properties driven by the emergent gauge field. In particular, while the Abelian gauge field coming from $4\pi$ vortices induces the two-dimensional topological Hall effect for a polarized magnon current, the non-Abelian gauge field generated by Shankar skyrmions gives rise to its three-dimensional counterpart. Our work offers concrete examples of non-Abelian interactions between magnetic solitons and excitations on top of them, therefore enriching and expanding the field of topological magnonics beyond the conventional Abelian paradigm.

\textit{Effective theory for magnons.}|Three-dimensional rotation matrices are efficiently parametrized by unit-norm quaternions, $R_{\alpha\beta}=(1-2\bm{v}^{2})\delta_{\alpha\beta}+2v_{\alpha}v_{\beta}-2\epsilon_{\alpha\beta\gamma}w v_{\gamma}$, where $\mathbf{q}\equiv(w,\bm{v})$ denotes the quaternion, $w^{2}+\bm{v}^{2}=1$ describes the unit-norm constraint~\cite{SM} and $\alpha,\beta,\gamma=x,y,z$ run over spatial indices. The quaternion representation of the action [Eq.~(\ref{eq:NsM_O4})] is given by $\mathcal{S}[\mathbf{q}]=\int dt\,d\vec{r}\,\left[\tfrac{1}{2}\partial_{\mu}\mathbf{q}\odot\partial^{\mu}\mathbf{q}\right]$, where $\odot$ denotes the scalar product of quaternions. To obtain the effective theory for small-amplitude fluctuations $\delta \mathbf{q} (\vec{r}, t)$ on top of a spatiotemporal quaternion texture $\mathbf{q}^0 (\vec{r}, t)$, it is convenient to work with the local quaternion frame within which the quaternion texture is uniform. To this end, we now consider a SO(4) rotation of the quaternion (understood as an element of $\mathbb{R}^{4}$), $\mathcal{R}$, that locally maps it onto a reference quaternion field $\mathbf{q}_{e}$, namely $\mathbf{q}(\vec{r},t)=\mathcal{R}(\vec{r},t)\mathbf{q}_{e}(\vec{r},t)$. Changing of the frame yields $\partial_{\mu}\mathbf{q}=\mathcal{R}\left[\partial_{\mu}+\mathcal{A}_{\mu}\right]\mathbf{q}_{e}$, where the matrices $\mathcal{A}_{\mu}\equiv\mathcal{R}^{\top}\partial_{\mu}\mathcal{R}$ encapsulate the spatiotemporal variations of the order parameter. To work within the local quaternion frame, it is natural to introduce a covariant derivative $D_{\mu}\equiv\partial_{\mu}+\mathcal{A}_{\mu}$, with $\mathcal{A}_{\mu}$ specifying the local connection for the geometry of the quaternion (spin) space. We note in passing that, from the orthogonality condition $\mathcal{R}^{\top}\mathcal{R}=\mathcal{R}\mathcal{R}^{\top}=\mathds{1}_{4}$, we conclude that the matrices $\mathcal{A}_{\mu}$ are antisymmetric. By invoking the invariance of  the inner-product operator $\odot$ under SO(4) rotations, the effective action in the quaternion representation can be recast as
\begin{equation}
\label{eq:NsM_versors2}
\mathcal{S}[\mathbf{q}_{e}]=\int dt\,d\vec{r}\,\left[\frac{1}{2}D_{\mu}\mathbf{q}_{e}\odot D^{\mu}\mathbf{q}_{e}\right].
\end{equation}

The order-parameter fluctuations are introduced into our effective model by superimposing SO(3) magnons on the uniform \emph{ground-state} reference quaternion field. We make the choice $\mathbf{q}^{0}_{e}\equiv(1,0,0,0)$ since it parametrizes the identity matrix $\mathds{1}_{3}$ and, therefore, SO(3) magnons, which correspond to small deviations around (and orthogonal to) $\mathbf{q}^{0}_{e}$, parametrize small-angle rotations generated via the exponential map. More specifically, we have $\mathbf{q}_{e}\equiv\mathbf{q}^{0}_{e}+\delta\bm{v}$, with $\delta\bm{v}\equiv(0,\delta v_{x},\delta v_{y},\delta v_{z})$, $||\delta\bm{v}||\ll1$. By introducing this quaternion decomposition into Eq.~\eqref{eq:NsM_versors2} and expanding it up to second order in the magnon field $\Psi\equiv\mathcal{P}^{\dagger}\delta\bm{v}=(\delta v_{x},\delta v_{y},\delta v_{z})$ with $\mathcal{P}^{\dagger}\equiv(\bm{0}|\mathds{1}_{3})$ a projection operator onto the magnon space, we obtain the following Lagrangian density for the SO(3) magnons~\cite{SM}:
\begin{equation}
\label{eq:Lag_SW3}
\mathcal{L}_{\textrm{SW}}\equiv\tfrac{1}{2}\underline{D}_{\mu}\Psi^{\dagger}\underline{D}^{\mu}\Psi,
\end{equation}
where  $\underline{D}_{\mu}=\partial_{\mu}+\underline{\mathcal{A}}_{\mu}$ is the new covariant derivative and $\underline{\mathcal{A}}_{\mu}\equiv\mathcal{P}^{\dagger}\mathcal{A}_{\mu}\mathcal{P}$ defines the new projected local connection for the quaternion geometry.

Since the gauge fields $\underline{\mathcal{A}}_{\mu}$ are antisymmetric, they can be identified with a gauge vector $\bm{\underline{A}}_{\mu}$ according to the expression $\underline{\mathcal{A}}_{\mu}=\underline{A}_{\mu}^{k}\hat{L}_{k}$, where $\{\hat{L}_{k}\}_{k=1}^{3}$ is the set of generators of three-dimensional rotations, i.e., the basis of the $\mathfrak{so}(3)$ Lie algebra, whose matrix coefficients read $\hat{L}_{k}|_{ij}=\epsilon_{kij}$, and hence $\underline{A}_{\mu}^{k}=- (1/2) \textrm{Tr}[\underline{\mathcal{A}}_{\mu}\hat{L}_{k}]$. We note in passing that SO(4) rotations that map $\mathbf{q}_{e}^{0}$ to the quaternion order parameter $\mathbf{q}$ are not uniquely determined. However, if $\mathbf{q}_{e}^{0}\neq\mathbf{q}$, there exists one and only SO(4) rotation $\mathcal{R}$ mapping $\mathbf{q}_{e}^{0}$ to $\mathbf{q}$ and leaving the orthogonal subspace $\langle\mathbf{q}_{e}^{0},\mathbf{q}\rangle^{\perp}$ invariant, and it is given by $\mathcal{R}=\mathds{1}-\frac{1}{1+\mathbf{q}_{e}^{0}\odot\mathbf{q}}(\mathbf{q}_{e}^{0}+\mathbf{q})\big[(\mathbf{q}_{e}^{0}+\mathbf{q})\odot\big]+2\mathbf{q}[\mathbf{q}_{e}^{0}\odot\hspace{0.1cm}]$. With this choice of SO(4) rotation matrix, the resultant local connection $\underline{\mathcal{A}}_{\mu}$ has the following unexpectedly compact form:
\begin{equation}
\label{eq:gauge_field}
\underline{\mathcal{A}}_{\mu}\big|_{\alpha\beta}=\frac{1}{1+w}(v_{\alpha}\partial_{\mu}v_{\beta}-v_{\beta}\partial_{\mu}v_{\alpha}) \, .
\end{equation}
The effective theory for SO(3) magnons on top of an arbitrary background texture [Eq.~(\ref{eq:Lag_SW3})] with the compact expression for the corresponding gauge field [Eq.~(\ref{eq:gauge_field})] is one of our main results. In what follows we will specify these gauge fields for the typical solitons emerging in the SO(3) order parameter, namely $4\pi$-vortices, Shankar skyrmions, and magnetic disclinations. These three types of solitons will be shown to exhibit an Abelian, non-Abelian, and vanishing topological electromagnetic field, respectively, on magnons, which will be followed by the discussion of the ensuing emergent Hall physics of SO(3) magnons.

\begin{figure}
\includegraphics[width=1.0\columnwidth]{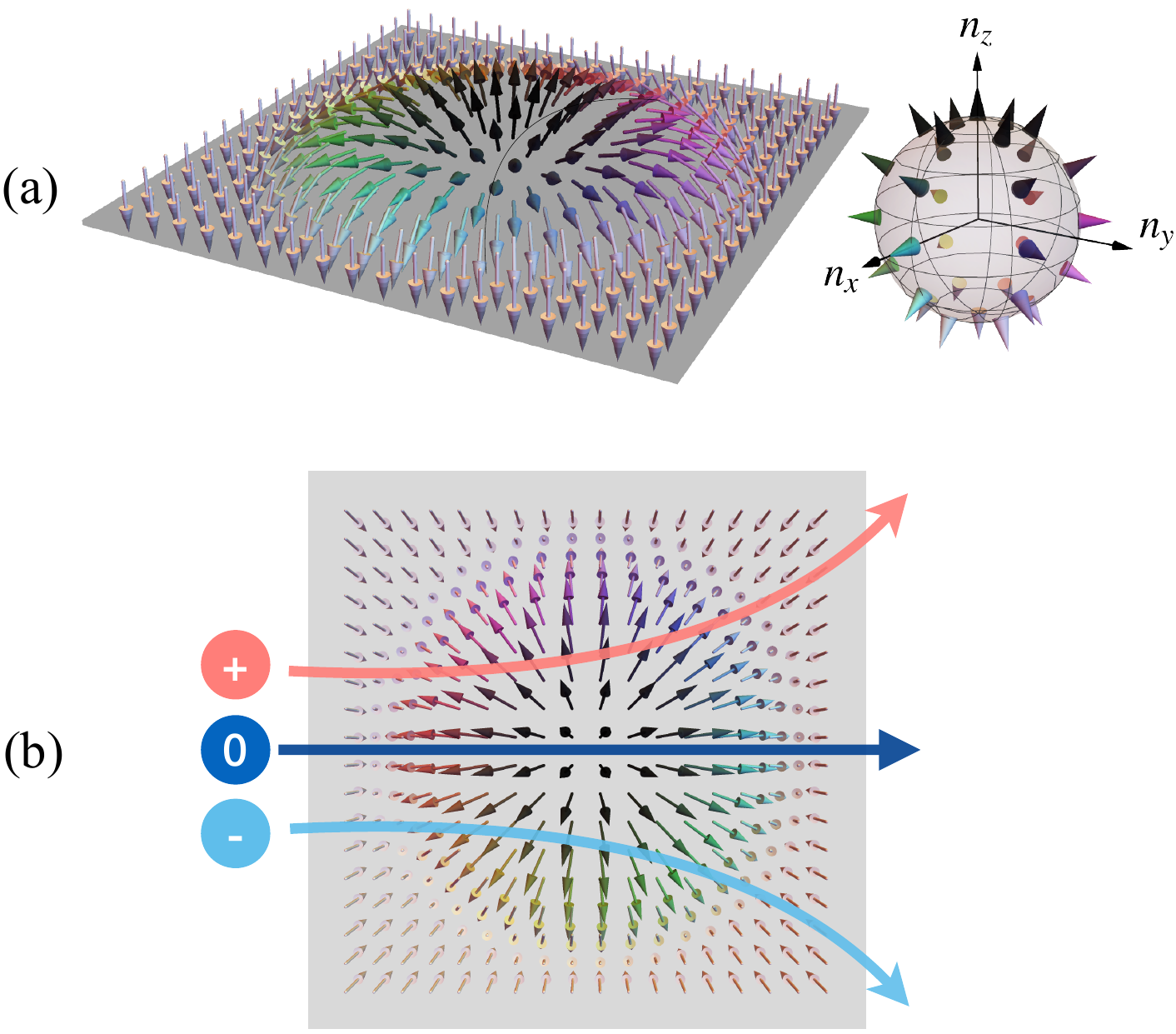}
\caption{(a) The spatial profile of the unit vector $\hat{R}(x,y)\hat{\mathbf{z}}$ for a 4$\pi$-vortex. The color scheme for the arrows is shown on the right. (b) The schematic illustration of the Hall effect on three types of magnons exerted by the emergent magnetic field induced by a 4$\pi$-vortex.}
\label{fig:4pi}
\end{figure}

\textit{$4\pi$-vortices.}|We begin by discussing two-dimensional SO(3) solitons, namely $4\pi$-vortex tubes also known as the Anderson-Toulouse vortices. We take the following ansatz for a $4\pi$-vortex in the quaternion parametrization~\cite{Anderson-PRL1976}
\begin{equation}
w(\vec{r}) = \cos\big[\tfrac{1}{2}f(\rho)\big] \, , \hspace{0.1cm}
v_{\alpha}(\vec{r}) = \sin\big[\tfrac{1}{2}f(\rho)\big]\hat{e}_{\phi}\cdot\hat{e}_{\alpha},
\end{equation}
where $(\rho,\phi)$ denote the radius and azimuthal angle in cylindrical coordinates, and $\hat{e}_{\rho}=(\cos\phi,\sin\phi,0)$ and $\hat{e}_{\phi}=(-\sin\phi,\cos\phi,0)$ are the unit radial and azimuthal vectors, respectively. Here, $f(\rho)$ is a monotonic function of $\rho$ parametrizing the angle of rotation induced by the Anderson-Toulouse vortex and satisfying the boundary conditions $f(0)=0$ and $f(R_{v})=\pi$ with $R_v$ the size of the soliton. The $z$-axis projection of the SO(3) order parameter $\hat{R}(\rho, \phi)$ corresponding to the quaternion order parameter is visualized in Fig.~\ref{fig:4pi}(a). The other projections of the SO(3) order parameter can be found in Ref.~\cite{SM}. The magnon gauge fields for this class of SO(3) solitons read~\cite{SM}
\begin{equation}
\underline{\mathcal{A}}_{\mu}^{\textrm{AT}}=(1-\delta_{\mu z})\big(1-\cos\big[\tfrac{1}{2}f(\rho)\big]\big)\partial_{\mu}\phi \hat{L}_{z} \, .
\end{equation}
These gauge fields turn out to be diagonalizable simultaneously, $\underline{\mathcal{A}}_{\mu}^{\textrm{AT}}\rightarrow i\underline{A}^{\textrm{AT}}|_{\mu}^{z}\textrm{diag}(1,0,-1)$ with $\underline{A}^{\textrm{AT}}|_{\mu}^{z}=(1-\delta_{\mu z})\big(1-\cos\big[\tfrac{1}{2}f(\rho)\big]\big)\partial_{\mu}\phi$. The change of basis for the magnon field to the chiral basis reads $\Psi=(\Psi_{+}, \Psi_{0}, \Psi_{-})^{\top}$, where $\Psi_{\pm}\equiv\tfrac{1}{\sqrt{2}}\left(\Psi_{x}\mp i\Psi_{y}\right)$ and $\Psi_{0}\equiv\Psi_{z}$.  In this basis, the effective magnon Lagrangian~\eqref{eq:Lag_SW3} is recast as
\begin{equation}
\label{eq:Lag_SW4}
\mathcal{L}_{\textrm{SW}}\equiv\tfrac{1}{2}\hspace{-0.25cm}\sum_{q=-1,0,1}\hspace{-0.25cm}\big(\partial_{\mu}-iq\underline{A}^{\textrm{AT}}|_{\mu}^{z}\big)\Psi_{q}^{\star}\big(\partial^{\mu}+iq\underline{A}^{\textrm{AT}}|^{z,\mu}\big)\Psi_{q},
\end{equation}
where `$^{\star}$' denotes complex conjugation. Remarkably, it splits into three decoupled copies of the Lagrangian describing the dynamics of a charged particle subjected to an external electromagnetic field (engendered by the Abelian gauge field $\overrightarrow{\underline{A}^{\textrm{AT}}|^{z}}$), where the charge is given by three flavors $q=+1,0,-1$. In particular, the local expression for the emergent magnetic field reads $\vec{B}_{t}^{\textrm{AT}}=\vec{\nabla}\times\overrightarrow{\underline{A}^{\textrm{AT}}|^{z}}=-\frac{1}{\rho}\frac{d}{d\rho}\cos\left[\tfrac{1}{2}f(\rho)\right]\hat{e}_{z}$. The trajectories of magnons are bent by the emergent magnetic field of the $4\pi$-vortex and the deflection direction depends on the charge $q$ of magnons as schematically shown in Fig.~\ref{fig:4pi}(b).

Within a semiclassical approach, namely magnons being represented by wave packets with well-defined center $\vec{r}_{c}$ and momentum $\vec{p}_{c}$ in the phase space, their dynamics are described by the Newton-like equation of motion $\dot{\vec{p}}_{c}=q\vec{E}_{t}^{\textrm{AT}}+q\,\dot{\vec{r}}_{c}\times\vec{B}_{t}^{\textrm{AT}}$~\cite{Xiao-RMP2010,FN1}, where $\vec{E}_{t}^{\textrm{AT}}$ is the emergent electric field from the gauge field $\overrightarrow{\underline{A}^{\textrm{AT}}}$. The change of the magnon momentum for the given time interval is therefore given by $\Delta p\equiv q\int dt \,\vec{E}_{t}^{\textrm{AT}}\big[\vec{r}_{c}(t),t\big]+q\int dt\,\dot{\vec{r}}_{c}(t)\times\vec{B}_{t}^{\textrm{AT}}\big[\vec{r}_{c}(t),t\big]$, which depends on the particular trajectory followed by each magnon. In what follows, we assume the ballistic motion of magnons over length scales larger than the typical size of the $4\pi$-vortices, i.e. $v_{s}\tau_{s}\gg R_{v}$, where $v_{s}$ denotes the spin-wave velocity of the magnetic medium and $\tau_{s}$ defines the magnon scattering time. From the hydrodynamic standpoint, the state of the three-flavored magnons is fully described by the pairs $\{n_{q},\vec{v}_{q}\}$, with $n_{q}$ and $\vec{v}_{q}$ being the \emph{charge-q} magnon density and velocity field, respectively. In turn, the charge-q magnon current and spin-polarized magnon current can be defined as $\vec{j}_{q}^{M}\equiv n_{q}\vec{v}_{q}$ and $\vec{j}_{q}^{s,M}\equiv qn_{q}\vec{v}_{q}=q\,\vec{j}_{q}^{M}$, respectively. By considering the Drude model for magnon scattering and the stationary regime, we obtain $\vec{v}_{q}=\vec{v}_{q,0}+\delta\vec{v}_{q}$, where $\vec{v}_{q,0}$ denotes the velocity field of the (injected) magnons in ballistic motion and $\delta\vec{v}_{q}=q \tau_{s}\big[\vec{E}_{t}^{\textrm{AT}}+\vec{v}_{q,0}\times\vec{B}_{t}^{\textrm{AT}}\big]$. As a result, the charge-q magnon current becomes $\vec{j}_{q}^{M}\equiv\vec{j}_{q,0}^{M}+n_{q}\delta\vec{v}_{q}=\vec{j}_{q,0}^{M}+q\tau_{s}\vec{j}_{q,0}^{M}\times\vec{B}_{t}^{\textrm{AT}}+qn_{q}\tau_{s}\vec{E}_{t}^{\textrm{AT}}$, with $\vec{j}_{q,0}^{M}=n_{q}\vec{v}_{q,0}$ being the injected charge-q magnon current. Finally, the total magnon current and the total spin-polarized current become
\begin{align}
\label{eq:total_magnon_currs}
\vec{j}^{M}&\equiv\vec{j}_{0}^{M}+\tau_{s}\vec{j}_{0}^{s,M}\times\vec{B}_{t}^{\textrm{AT}}+\tau_{s}\left(n_{+}-n_{-}\right)\vec{E}_{t}^{\textrm{AT}},\\
\vec{j}^{s,M}&\equiv\vec{j}_{0}^{s,M}+\tau_{s}\big(\vec{j}_{0}^{M}-\vec{j}_{0,0}^{M}\big)\times\vec{B}_{t}^{\textrm{AT}}+\tau_{s}(n_{+}+n_{-})\vec{E}_{t}^{\textrm{AT}},\nonumber
\end{align}
where $\vec{j}_{0}^{M}\equiv\vec{j}_{0,0}^{M}+\vec{j}_{+,0}^{M}+\vec{j}_{-,0}^{M}$ and $\vec{j}_{0}^{s,M}\equiv\vec{j}_{+,0}^{M}-\vec{j}_{-,0}^{M}$ are the total injected magnon current and total injected spin-polarized magnon current, respectively.

In the case of both magnon chiralities being equally populated ($n_{-}=n_{+}$) as well as being injected with the same current ($\vec{j}_{+,0}^{M}=\vec{j}_{-,0}^{M}$), we derive the expressions $\vec{j}^{M}=\vec{j}_{0}^{M}$ and $\vec{j}^{s,M}=2\tau_{s}\vec{j}_{+,0}^{M}\times\vec{B}_{t}^{\textrm{AT}}+2\tau_{s}n_{+}\vec{E}_{t}^{\textrm{AT}}$ for the aforementioned currents. Therefore, there is no net transverse magnon current but there is a Hall contribution to the spin-polarized magnon current. We note that, in the ballistic regime, the volume average of the transverse magnon spin-polarized current is parametrized by the volume average of the topological magnetic field, which reads $\langle \vec{B}^{\textrm{AT}}_{t}\rangle_{V}\equiv\frac{1}{\pi R_{v}^{2}}\int_{0}^{R_{v}}\rho d\rho\int_{0}^{2\pi}d\phi\,\vec{B}^{\textrm{AT}}_{t}=\frac{2}{R_{v}^{2}}\hat{e}_{z}$. This leads us to one of our main results: A $4\pi$-vortex in frustrated magnets will give rise to a Hall effect on the polarized magnon current via the emergent Abelian magnetic field.

\begin{figure}
\includegraphics[width=0.8\columnwidth]{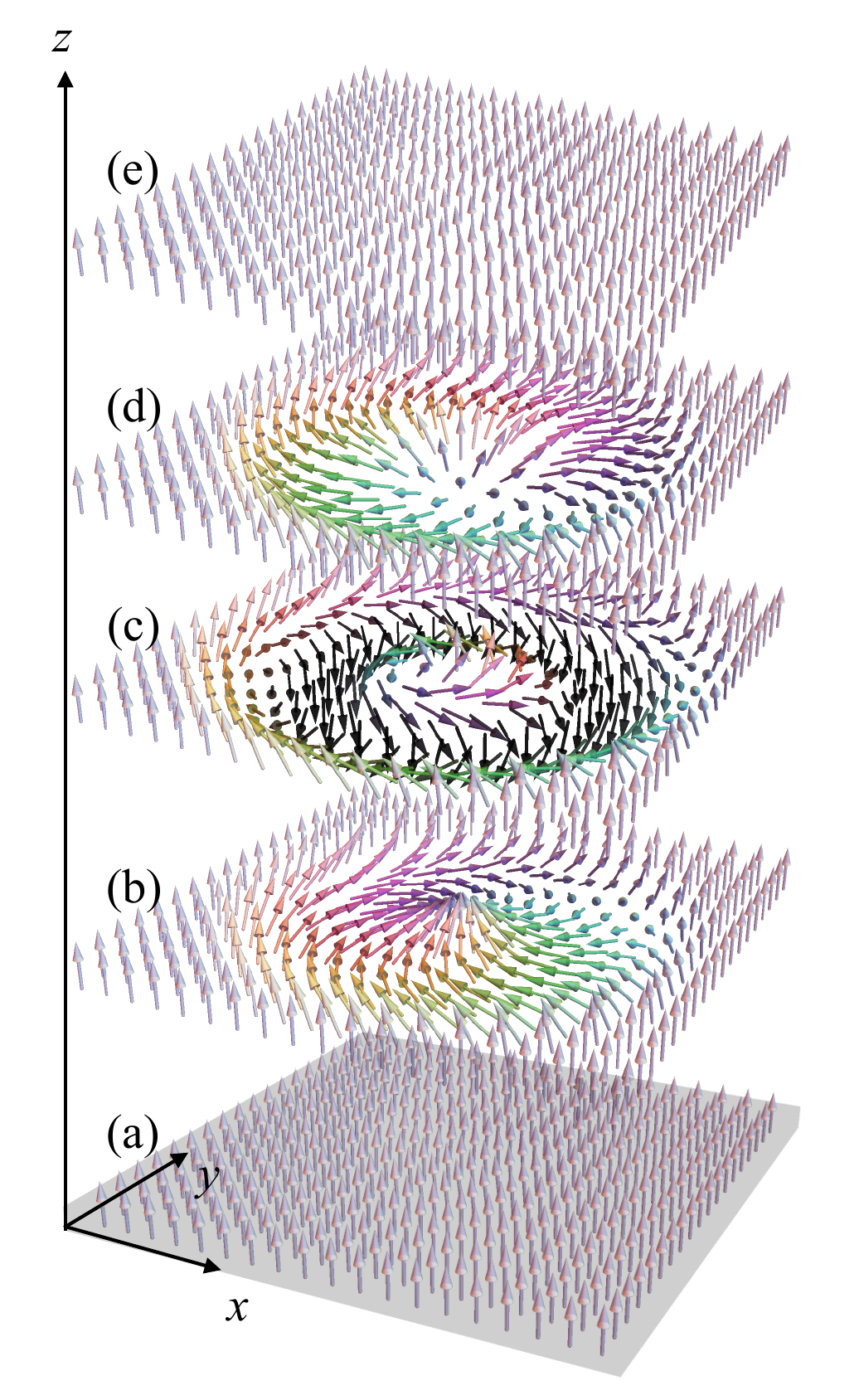}
\caption{The spatial profile of the unit vector $\hat{R}(x, y, z)\hat{\mathbf{z}}$ for a Shankar skyrmion for the fixed $z$-coordinates, (a) $z = - R_\text{Sk}$, (b) $z = -R_\text{Sk}/2$, (c) $z = 0$, (d) $z = R_\text{Sk}/2$, and (e) $z = R_\text{Sk}$. The color scheme for the arrows is the same as the one used for Fig.~\ref{fig:4pi}(a).}
\label{fig:shankar}
\end{figure}

\textit{Shankar skyrmions.}|Let us now turn to the second type of solitons, the so-called Shankar skyrmions, which are intrinsically three dimensional and classified by the (integer) $\pi_{3}$-homotopy invariant $\mathcal{Q}_{\textrm{sky}}\equiv\int d\vec{r}\left[\frac{1}{12\pi^{2}}\epsilon_{\alpha\beta\gamma}\epsilon_{klmn}q^{k}\partial_{\alpha}q^{l}\partial_{\beta}q^{m}\partial_{\gamma}q^{n}\right]$, where $\epsilon_{\iota_{1}\ldots\iota_{n}}$ denotes the Levi-Civita symbol~\cite{Zarzuela-PRB2019}. We take the following ansatz in the quaternion parametrization~\cite{Shankar-JPhys1977,Volovik-JETP1977}
\begin{equation}
\label{eq:v-sk}
w(\vec{r}) = \cos\big[\tfrac{1}{2}f(r)\big] \, , \hspace{0.1cm}
v_{\alpha}(\vec{r}) = \sin\big[\tfrac{1}{2}f(r)\big]\frac{x_{\alpha}}{r},
\end{equation}
where $f(r)$ is a radial monotonic function parametrizing the angle of rotation induced by the Shankar skyrmion as well as satisfying the boundary conditions $f(0)=2\pi$ and $f(R_\text{Sk})=0$ with $R_\text{Sk}$ the size of the soliton. The $z$-axis projection of the SO(3) order parameter $\hat{R}(x,y,z)$ for the Shankar skrymions is shown in Fig.~\ref{fig:shankar}. It is noteworthy that the two-dimensional spin configuration at $z = 0$ [Fig.~\ref{fig:shankar}(c)] is a so-called skyrmionium that is a topological spin texture consisting of two skyrmions with opposite topological charges adding up to zero. For the other projections of the SO(3) order parameter can be found in Ref.~\cite{SM}. The resultant skyrmion charge reads $\mathcal{Q}_{\textrm{sky}}=-1$. The matrix coefficients of the gauge fields for the Shankar skyrmion read 
\begin{equation}
\underline{\mathcal{A}}_{\mu}^{\textrm{Sk}}\big|_{\alpha\beta}=\frac{1-\cos[\tfrac{1}{2}f(r)]}{r^{2}}(x_{\alpha}\delta_{\mu\beta}-x_{\beta}\delta_{\mu\alpha}) \, ,
\end{equation}
so that $\{\underline{\mathcal{A}}_{x}^{\textrm{Sk}},\underline{\mathcal{A}}_{y}^{\textrm{Sk}},\underline{\mathcal{A}}_{z}^{\textrm{Sk}}\}$ do not commute among themselves. Thus, unlike the case of $4\pi$-vortices, we cannot introduce a new basis for the magnon field in which, from the perspective of the effective Lagrangian~\eqref{eq:Lag_SW3}, the projections of $\Psi$ are labelled by integer multiples of a quantum of topological charge. We conclude that the topological electromagnetic field engendered by this class of SO(3) solitons is purely non-Abelian. The parent Faraday tensor for the gauge field can be obtained from the commutator of the covariant derivatives $[\underline{D}_{\mu}^{\textrm{Sk}},\underline{D}_{\nu}^{\textrm{Sk}}]\equiv f_{\mu\nu}^{\textrm{Sk},k}\hat{L}_{k}$, namely
\begin{equation}
\label{eq:Shankar_Faraday}
\bm{f}_{\mu\nu}^{\textrm{Sk}}=\partial_{\mu}\underline{\bm{A}}_{\nu}^{\textrm{Sk}}-\partial_{\nu}\underline{\bm{A}}_{\mu}^{\textrm{Sk}}-\underline{\bm{A}}_{\mu}^{\textrm{Sk}}\otimes\underline{\bm{A}}_{\nu}^{\textrm{Sk}} \, ,
\end{equation}
where the last term is finite and thereby showing the non-Abelian nature of the gauge field. A semiclassical description of the magnon dynamics can be developed in the non-Abelian case, akin to that introduced in Ref.~\cite{Zarzuela-PRB2023} in the context of magnetically frustrated conductors. From Eq.~\eqref{eq:Lag_SW3} we can identify the magnon Hamiltonian as $\mathcal{H}_{\textrm{SW}}=\frac{1}{2}\underline{D}_{k}\Psi^{\dagger}\underline{D}_{k}\Psi+\ldots$, where $'\ldots'$ denotes other terms irrelevant for our discussion. Partial integration yields the following (quantum-mechanical) operator expression, $\hat{\mathcal{H}}_{\textrm{SW}}=\tfrac{1}{2}\big[\vec{p}-i\vec{\underline{\mathcal{A}}}\big]^{2}$. Within the Heisenberg picture, the velocity operator takes the form $\tfrac{d x_{i}}{dt}=-i\big[x_{i},\hat{\mathcal{H}}_{\textrm{SW}}\big]=\vec{p}-i\vec{\underline{\mathcal{A}}}$, namely $\vec{\Pi}\equiv\vec{p}-i\vec{\underline{\mathcal{A}}}$ defines the kinetic momentum of magnons. In turn, since $\hat{\mathcal{H}}_{\textrm{SW}}=\tfrac{1}{2}\vec{\Pi}^{2}$, the acceleration operator becomes~\cite{SM}
\begin{equation}
\label{eq:accelerator}
\frac{d\,\vec{\Pi}}{dt}=-i\big[\vec{\Pi},\hat{\mathcal{H}}_{\textrm{SW}}\big]=\frac{i}{2}\left[\vec{\Pi}\times\big(\vec{\bm{B}_{t}}\circ\bm{L}\big)-\big(\vec{\bm{B}_{t}}\circ\bm{L}\big)\times\vec{\Pi}\right],
\end{equation}
where $\vec{\bm{B}_{t}}$ is the topological magnetic field engendered by the texture of the SO(3) order parameter. We note that it reads $\bm{B}_{t}|_{k}=\tfrac{1}{2}\epsilon_{kij}\bm{f}_{ij}$ in terms of the Faraday tensor and that $\vec{\bm{B}_{t}}\circ\bm{L}\equiv B_{t}|^{k}_{j}\hat{L}_{k}\hat{e}_{j}$ denotes its contraction to the vector of generators of the $\mathfrak{so}(3)$ Lie algebra in the spin space. Here, we have invoked the nontrivial commutator relation $[\Pi_{i},\Pi_{j}]=-\epsilon_{ijk}\big(\bm{B}_{t}\circ\bm{L}\big)|_{k}$ for the components of the kinetic momentum. The average value of the acceleration operator is obtained by tracing Eq.~\eqref{eq:accelerator} with the density matrix of the magnon ensemble, $\textrm{Tr}\big[\hat{\rho}\ldots\big]$. Under the assumption of an external-force-driven magnon motion and the steady state, we can take $\vec{\Pi}_{0}\equiv\vec{\textrm{cnt}}$ for the leading (convective) contribution to the kinetic momentum and, therefore, 
\begin{equation}
\label{eq:accel_ave}
\Big\langle\frac{d\,\vec{\Pi}}{dt}\Big\rangle=\vec{\Pi}_{0}\times\big(\vec{\bm{B}_{t}}\circ\bm{p}\big), \hspace{0.5cm}\bm{p}\equiv \textrm{Tr}\big[i\hat{\rho}\,\bm{L}\big].
\end{equation}
Under the assumption of the ballistic motion of SO(3) magnons over length scales larger than the typical size of Shankar skyrmions, i.e. $v_{s}\tau_{s}\gg R_{\textrm{Sk}}$, the quantity parametrizing the magnon Hall dynamics is the volume average of the topological magnetic field, $\langle \bm{B}^{\textrm{Sk}}_{t}|_{k}\rangle\equiv\tfrac{1}{V_{\textrm{Sk}}}\int_{0}^{R_{\textrm{Sk}}}r^{2}dr\int_{S^{2}}d\Omega\,\bm{B}^{\textrm{Sk}}_{t}|_{k}$. A direct calculation yields the expression $\big\langle B^{\textrm{Sk}}_{t}|_{k}^{k}\big\rangle=-\frac{3}{2}\frac{1}{R_{\textrm{Sk}}^{2}},\hspace{0.2cm}\big\langle B^{\textrm{Sk}}_{t}|_{k}^{j}\big\rangle=0$ otherwise, for its matrix components~\cite{SM}, which indicates that the polarized magnon current experiences the topological emergent magnetic field applied in the same direction as the polarization. This leads us to one of our main results: A Shankar skyrmion gives rise to the Hall effect for a polarized magnon current and its deflection direction depends on the polarization due to its non-Abelian nature. Here, we remark that the polarization direction and the field direction are the same in our case since the emergent magnetic-field tensor $\vec{\bm{B}_{t}}$ is diagonal, but they differ in general cases. In addition, by invoking the argument analogous to the $4\pi$-vortex case, we predict that the injection of a unpolarized magnon current along the $x$ direction will induce the $z$-polarized magnon current in the $y$ direction and $y$-polarized magnon current in the $z$ direction, respectively. The cases for the unpolarized magnon current in the $y$ and $z$ directions can be obtained with cyclic permutation of $x, y, z$.

\textit{Discussion.}|In addition to the aforementioned topological solitons, there is a topological defect associated with the SO(3) order parameter, the so-called magnetic disclination, that carries a Z$_2$ topological charge. In the case of straight magnetic disclinations, which are amenable to be engineered in a realistic frustrated magnet~\cite{Zarzuela-PRB2023}, we obtain the identity $\underline{\mathcal{A}}_{\mu}^{\textrm{md}}=\bm{0}_{4\times4}$~\cite{SM}. Therefore, this class of topological singularities does not engender an effective electromagnetic response for the magnon dynamics.

The non-Abelian semiclassical description for three-flavored magnon dynamics, encapsulated in Eq.~\eqref{eq:accel_ave}, does also apply to the case of $4\pi$-vortices and yields the same results discussed before: the gauge vector takes the form $\bm{A}_{\mu}^{\textrm{AT}}=\underline{A}^{\textrm{AT}}|^{z}_{\mu}\bm{\hat{e}}_{z}$ in this case, so that the corresponding Faraday tensor reads $\bm{f}_{\mu\nu}^{\textrm{AT}}=\big(\partial_{\mu}\underline{A}^{\textrm{AT}}|^{z}_{\nu}-\partial_{\nu}\underline{A}^{\textrm{AT}}|^{z}_{\mu}\big)\bm{\hat{e}}_{z}$ and, thus, reduces to the expression expected in the Abelian scenario. As a result, the contraction $\vec{\bm{B}_{t}}\circ\bm{p}$ in Eq~\eqref{eq:accel_ave} becomes $p_{z}\vec{B}_{t}^{\textrm{AT}}$, which implies that magnons polarized within the basal plane of the $4\pi$-vortex do not experience any Hall deflection during their motion. In this regard, the polarization vector ascribed to the three different chiral magnon modes can be calculated as $\bm{p}|_{q}=\langle\Psi_{q}|i\bm{L}|\Psi_{q}\rangle$, $q=0,\pm$. A direct calculation yields the values $\bm{p}|_{\pm}=\pm \bm{\hat{e}}_{z}$ and $\bm{p}|_{0}=\bm{0}$ for the magnon polarization, from which we conclude again that the chiral magnon modes $q=0,\pm$ carry spin $0,\pm \hbar$, respectively, along the direction of the $4\pi$-vortex axis~\cite{SM}.

Our magnon gauge field $\underline{\mathcal{A}}_{\mu}$ is real and antisymmetric and therefore is of SO(3) nature. It is an open question if there is a spin system where fluctuations on top of a texture are ought to be described by SU(3) rather than SO(3). If the answer is positive, the corresponding fluctuations will bear close resemblance to the quantum chromodynamics for quarks, where the gluon field is described by the SU(3) non-Abelian gauge field. We speculate that inclusion of the certain symmetry-breaking effects such as the Dzyaloshinskii-Moriya interaction for the inversion symmetry and the dissipation for the time reversal symmetry may offer a way to realize the SU(3) magnon gauge field by bringing about the chiral and the non-Hermitian effects into the magnon system. In addition, we have adopted three-dimensional rotation matrices and the corresponding quaternions as the order parameter of frustrated magnets. An alternative ``spin-frame'' order-parameter representation, which is given by a orthogonal triplet of unit vectors, has recently been proposed in Ref.~\onlinecite{Pradenas-PRL2024}. It would be interesting to investigate how the non-Abelian gauge fields that we discussed here are manifested in the spin-frame field theory.

Lastly, for the experimental proposal, any magnetic material that harbors frustrated or non-collinear spin texture in equilibrium, e.g., spin glasses~\cite{Binder-RMP1986}, the correlated spin glass phase of amorphous magnets~\cite{Chudnovsky-PRB1983, Chudnovsky-PRB1986} and frustrated antiferromagnets~\cite{Dombre-PR1989, Azaria-PRL1992, Chubukov-PRL1994, Mook-PRB2017} can be a material platform. A polarized magnon current can be injected into a frustrated magnet by utilizing a heavy metal via the spin Hall effect as shown in Refs.~\onlinecite{Tserkovnyak-PRB2017, Ochoa-PRB2018}. After scattering with a $4\pi$-vortex or a Shankar skyrmion, the magnon current is expected to exhibit a topological Hall effect, which is non-Abelian in the latter case, and possess a transverse component, which can be detected by the inverse spin Hall effect. In addition, applying a temperature gradient in the considered setup will give rise to the spin Nernst effect~\cite{Meyer-NM2017}, similar to the antiferromagnetic counterparts~\cite{Diaz-PRL2019}.

\begin{acknowledgments} R.Z. was supported by the Deutsche Forschungsgemeinschaft (DFG, German Research Foundation) - SPP 2137 Skyrmionics (project 462597720) and the Dynamics and Topology Centre TopDyn funded by the State of Rhineland Palatinate. S.K.K. was supported by Brain Pool Plus Program through the National Research Foundation of Korea funded by the Ministry of Science and ICT (NRF-2020H1D3A2A03099291), by the National Research Foundation of Korea (NRF) grant funded by the Korea government (MSIT) (NRF-2021R1C1C1006273), and by the National Research Foundation of Korea funded by the Korea Government via the SRC Center for Quantum Coherence in Condensed Matter (NRF-RS-2023-00207732).
\end{acknowledgments}


\begin{thebibliography}{99}
\bibitem{FN1} We have assumed in this work the convention $\hbar\equiv1$.
\end{thebibliography}
\end{document}